\documentclass{emulateapj}

\usepackage{natbib}    
\usepackage{graphicx}

\begin{document}

\title{Identifying High Metallicity M Giants at Intragroup Distances with SDSS}

\author{Lauren E. Palladino,\altaffilmark{1} Kelly Holley-Bockelmann,\altaffilmark{1,2} Heather Morrison,\altaffilmark{3} Patrick R. Durrell,\altaffilmark{4}  Robin Ciardullo,\altaffilmark{5,6} John Feldmeier,\altaffilmark{4} Richard A. Wade,\altaffilmark{5} J. Davy Kirkpatrick,\altaffilmark{7} Patrick Lowrance\altaffilmark{7}}

 \affil{
     $^{1}$ Department of Physics and Astronomy, Vanderbilt University, Nashville, TN 37235\\
     $^{2}$ Department of Natural Sciences and Mathematics, Fisk University, Nashville, TN 37208\\
     $^{3}$ Department of Astronomy, Case Western Reserve University, Cleveland, OH 44106\\
     $^{4}$ Department of Physics and Astronomy, Youngstown State University, Youngstown, OH 44555 \\
     $^{5}$ Department of Astronomy and Astrophysics, The Pennsylvania State University, University Park, PA 16802\\
     $^{5}$ Institute for Gravitation and the Cosmos, The Pennsylvania State University, University Park, PA 16802, USA\\
     $^{7}$ Infrared Processing and Analysis Center, MS 100-22, California Institute of Technology, Pasadena, CA 91125\\
     lauren.e.palladino.1@vanderbilt.edu, k.holley@vanderbilt.edu}

\begin{abstract} 
Tidal stripping and three-body interactions with the central supermassive black hole may eject stars from the Milky Way. These stars would comprise a set of `intragroup' stars that trace the past history of interactions in our galactic neighborhood. Using the Sloan Digital Sky Survey DR7, we identify candidate solar metallicity red giant intragroup stars using color cuts that are designed to exclude nearby M and L dwarfs. We present 677 intragroup candidates that are selected between 300 kpc and 2 Mpc, and are either the reddest intragroup candidates (M7-M10) or are L dwarfs at larger distances than previously detected.
\end{abstract}

\section{INTRODUCTION}

A significant fraction of the stellar component of a galaxy cluster is not confined to any galaxy. These stars between galaxies form luminous halos, called intracluster light (ICL), with very low surface brightness that can extend out to several hundred kiloparsecs around individual galaxies \citep[eg.,][]{Abadi,Krick2007}. The brightest ICL is less than 1\% of the brightness of the night sky \citep{Mihos2003,Feldmeier2003}, thus making a complete census of ICL very difficult to obtain. High resolution N-body simulations estimate that ICL could comprise 10\%-70\% of the total cluster luminosity \citep{Mihos2003,Murante2004}. 

It is commonly thought that intracluster stars are caused by one of three main channels: 1) stripping from galaxies as the cluster assembles either via high speed galaxy encounters, tidal shocking, or a rapidly changing galaxy cluster potential \citep[][]{Byrd,Merritt}, 2) long-lived, low level cluster perturbations in the form of ``galaxy harassment" \citep{Moore}, or 3) tidal stripping within in-falling galaxy groups \citep{Mihos2004,Rudick}. These processes will generate a stellar `debris field' that is highly inhomogeneous, with distinctly non-Gaussian velocities that reflect an unrelaxed intracluster population \citep{Napolitano}. Thus far, ICL has been identified via planetary nebulae (PNe) \citep{Feldmeier2003,Aguerri}, Red Giant Branch stars \citep{Durrell2002,Williams}, intracluster globular clusters \citep{Lee2010,Peng2011}, and ultra-deep surface photometry \citep{Feldmeier2002,Feldmeier2004,Mihos2005}.

\begin{figure*}
\centering
\begin{tabular}{cc}
\includegraphics[width=3.25in,height=2.5in]{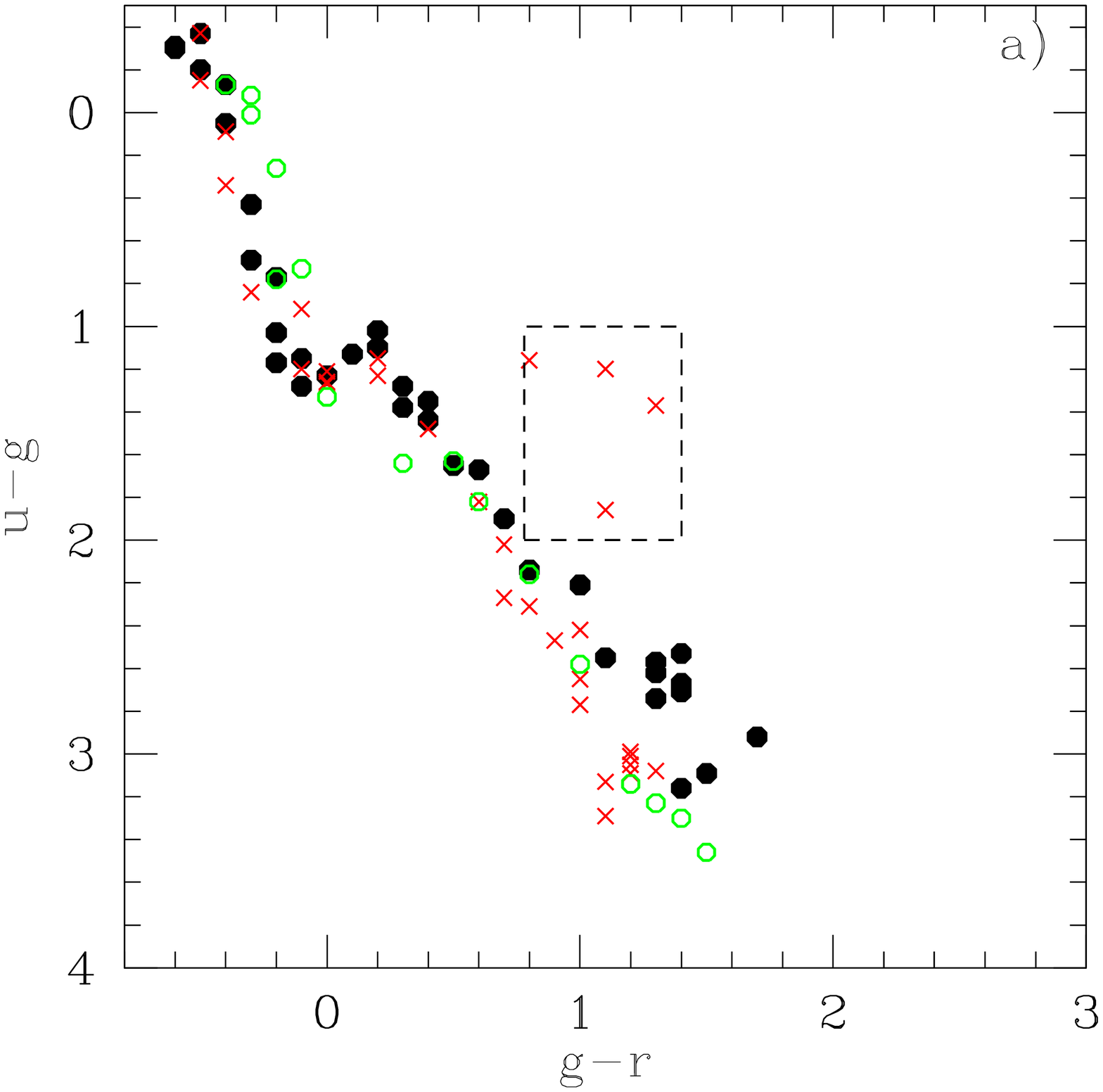} & \includegraphics[width=3.25in,height=2.5in]{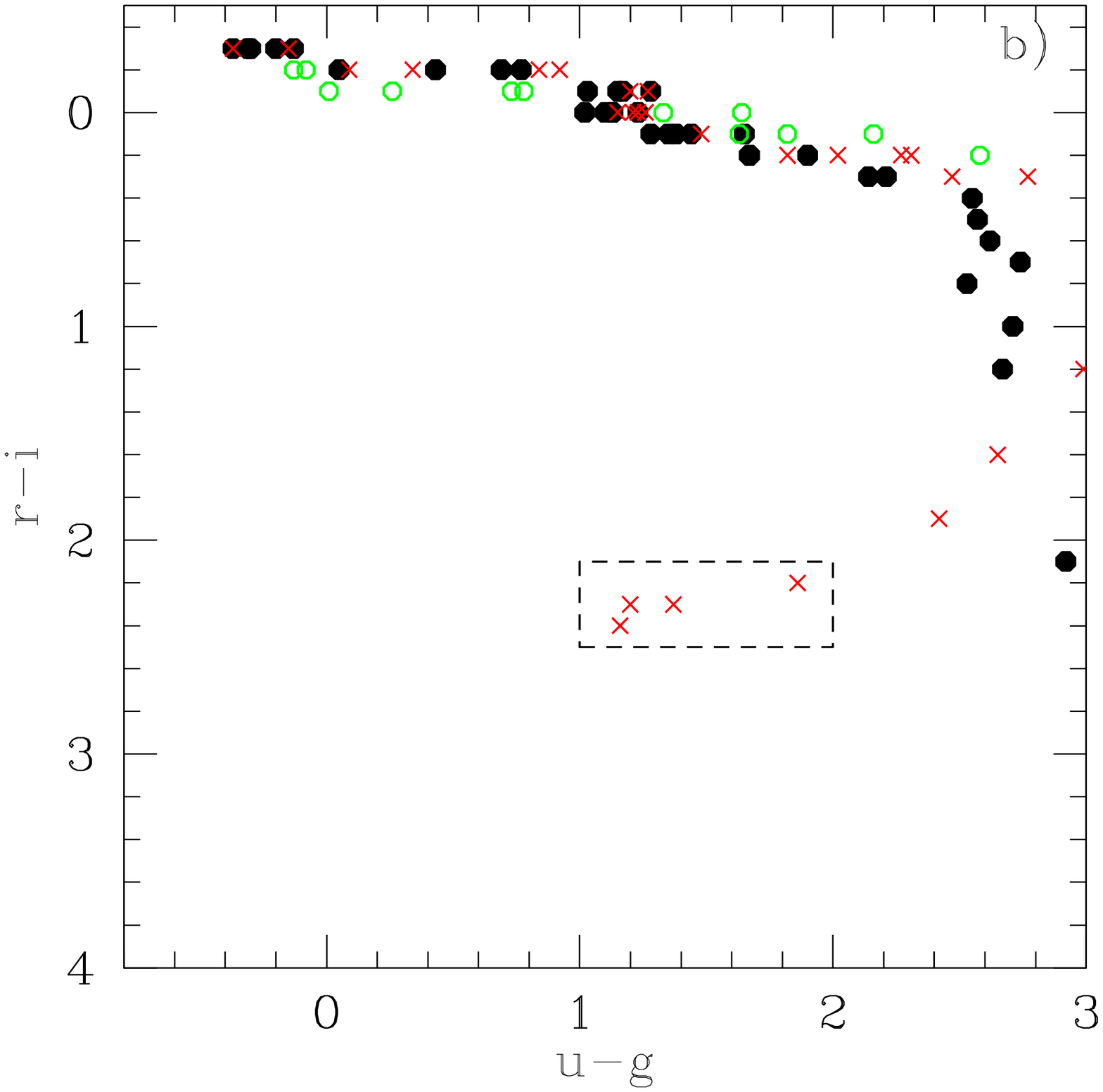} \\
\includegraphics[width=3.25in,height=2.5in]{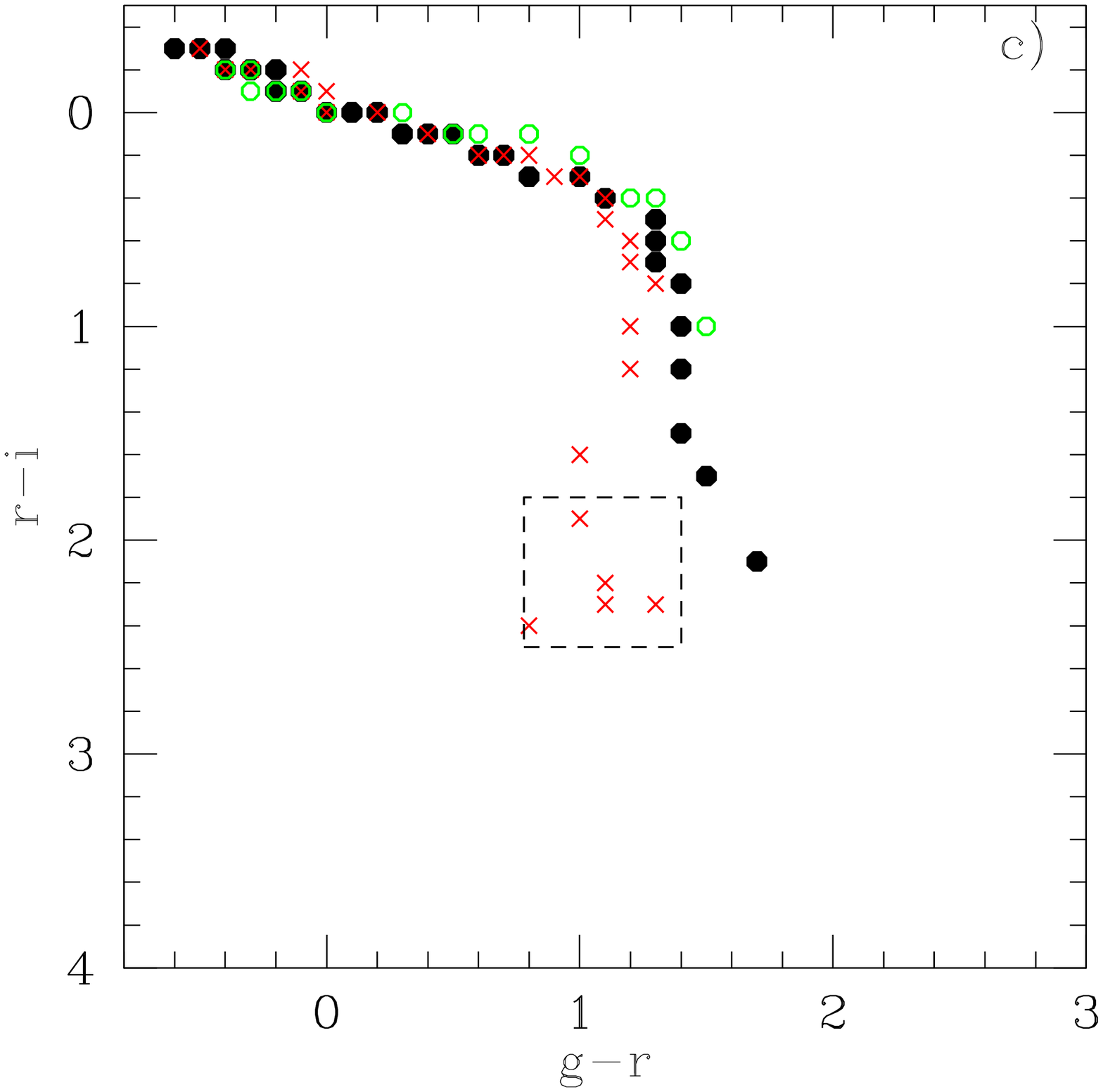} & \includegraphics[width=3.25in,height=2.5in]{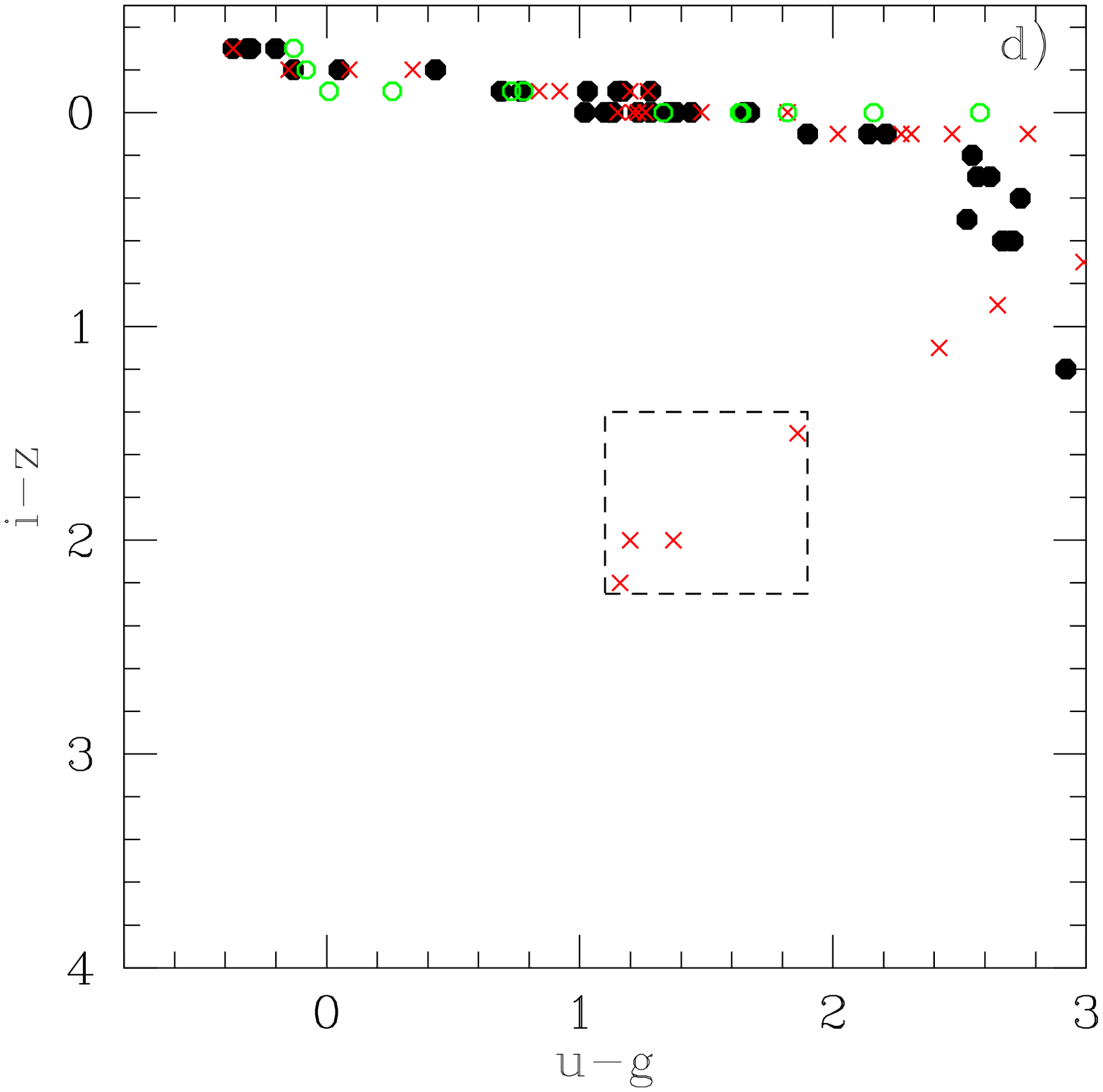} \\
\includegraphics[width=3.25in,height=2.5in]{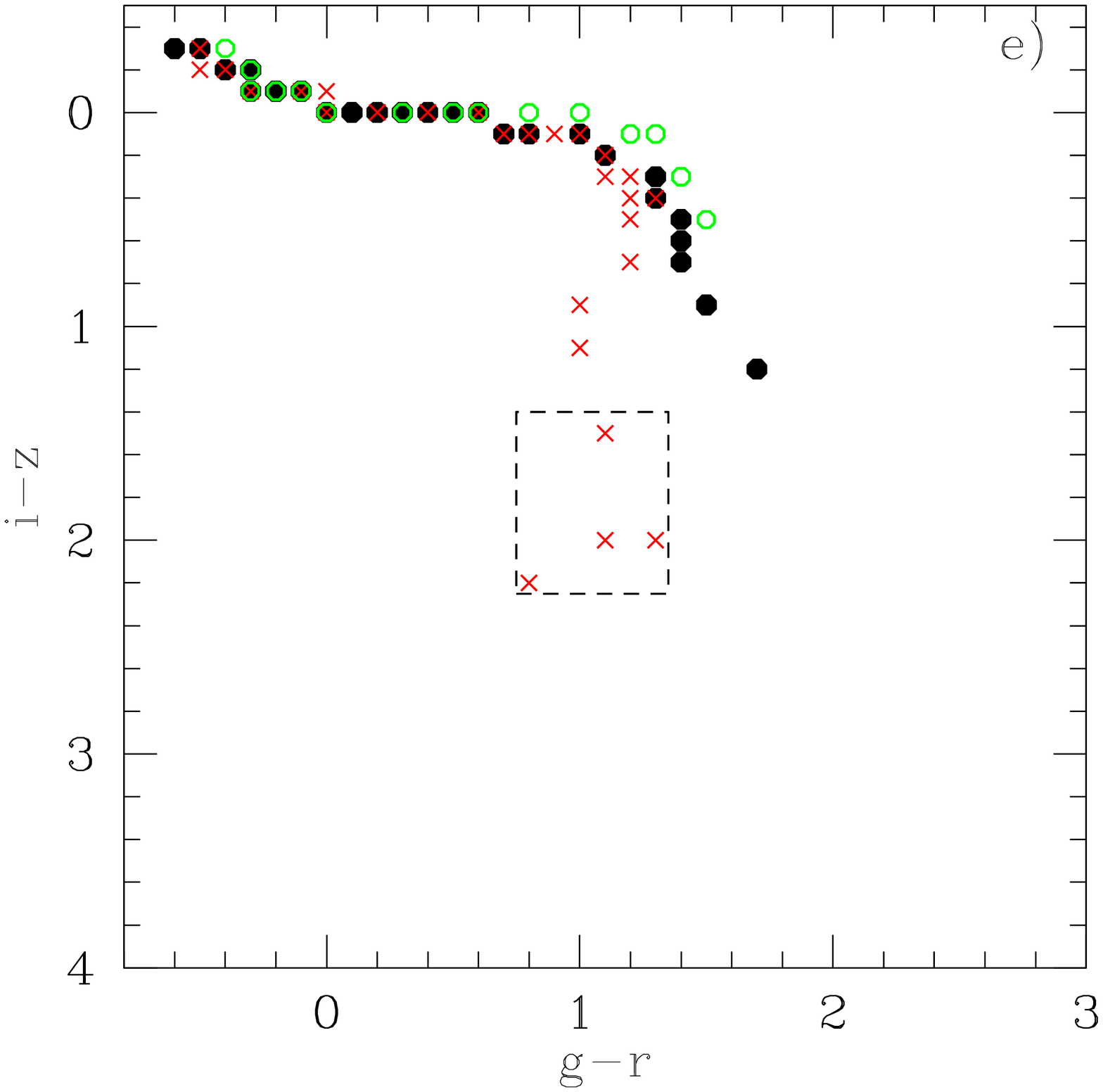} & \includegraphics[width=3.25in,height=2.5in]{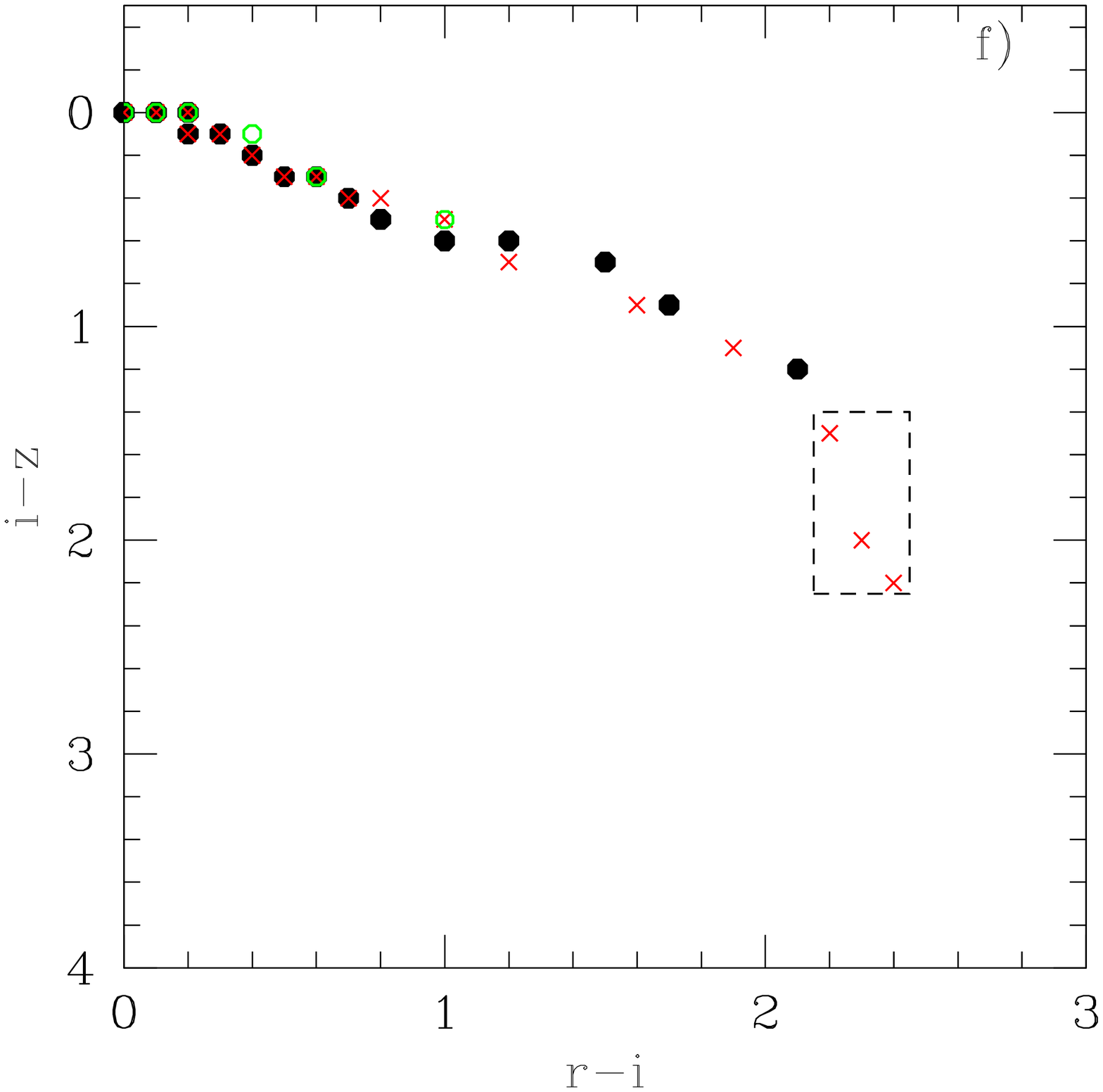} \\
\end{tabular}
\caption{Color-color diagrams of the \protect\cite{Covey} sample data. Main sequence stars are marked with black dots, giants with red crosses, and supergiants with green circles. Note that the reddest giants (M7-M10), identified by the black dashed regions, are isolated from the stellar locus.  This figure is a theoretical demonstration that very late M giants have the potential to be isolated in color space in the $\it{ugriz}$ bands, and these dashed regions are meant only to highlight the spectral types of interest and do not represent our color selection criteria (given explicitly in equations 1 and 2). In particular, we drop the $\it{u-g}$ and $\it{g-r}$ cuts because objects at these distances are likely too faint in the $\it{u}$ and $\it{g}$ bands. The locus of sub-solar metallicity giants is generally indistinguishable from the dwarf locus, and thus not plotted here.}
\label{}
\end{figure*}

The properties of ICL may provide insights into the accretion history and evolution of galaxy clusters \citep{Mihos2003,Mihos2004,Napolitano,Feldmeier2003,Conroy2007}. Although there is some debate about the role that tidal stripping plays in ICL production, it is expected that ICL substructure is correlated with the dynamical state of the cluster \citep{Murante2007,Mihos2004}. Since the vast majority of galaxies reside in poor groups, rather than in large clusters, it is of great interest to determine the fraction of unbound stars that reside in these environments.

In the Local Group, ICL has not been observed, though deep observations and star counts have revealed a ``field of streams" \citep[e.g.,][]{Belokurov}. These streams have been detected out to 100 kpc and are bound to the Milky Way  \citep[e.g.,][]{Yanny,Ibata}. Similarly, faint streams have been detected in the outskirts of M31 \citep{McConnachie,Ibata2007}. Given that the Milky Way and M31 are not yet interacting and may not even be part of the same dark matter halo, it is more likely that Local Group ICL, if it exists, would be a product of a different process altogether. 

One of the more recent suggestions for ICL production is via three-body interactions \citep{KHB}.
For example, stars can be thrown out from the galaxy through tidal disruption of a binary star system by a supermassive black hole \citep{Hills,Yu}; this is the most common explanation for `hypervelocity' stars such as SDSS J090745.0+0204507, with a galactic rest frame velocity of $\sim$ 700 km/sec \citep{Brown}. During this process, energy and angular momentum are transferred from the black hole to one of the stars in the binary.  The second star loses energy and becomes bound to the black hole while the first  is ejected from the galaxy. This is expected to occur at a rate of 
$10^{-5}(\eta /0.1)\, {\rm yr}^{-1},$
where $\eta$ is the stellar binary fraction \citep{Magorrian}.

Another three-body interaction that is likely to expel stars is a close encounter of a single star with a binary black hole \citep{Yu}. This is expected to occur at a rate of
$10^{-4}(\eta /0.1)\, {\rm yr}^{-1} $
\citep{Magorrian}. 
In this case, the star gains energy from the binary black hole and is flung out of the galaxy while the black hole orbit shrinks \citep[e.g.,][]{Quinlan,Sesana}. 

To become gravitationally unbound, stars must exceed the escape velocity of the Galaxy, now estimated to be 500-600 km/sec \citep[e.g.,][]{Smith}. Semi-analytic models predict that there may be approximately 100 hypervelocity stars within 8 kpc of the galactic center if the binary stars have equal masses \citep{Yu}. However, intragroup stars (IGS) may not be solely comprised of hypervelocity stars; they may still be bound but on very large, highly eccentric orbits-- this can increase the potential number of IGS. One way to get stars on such eccentric orbits is through three-body galaxy ejections of satellites like Leo I \citep{Sales,Mateo2008}.

As a first attempt to probe for a population of intragroup stars, we develop a technique to search for M giant stars in between the Local Group galaxies. In this paper we present our technique for identifying candidate IGS from the Sloan Digital Sky Survey (SDSS) by applying color, distance, and proper motion cuts. Section 2 describes our technique. We present our results in Section 3, and we discuss possible sources of contamination in Section 4. Section 5 concludes and discusses methods to confirm the candidates.

\begin{figure*}
\centering
\begin{tabular}{cc}
\includegraphics[width=3.2in]{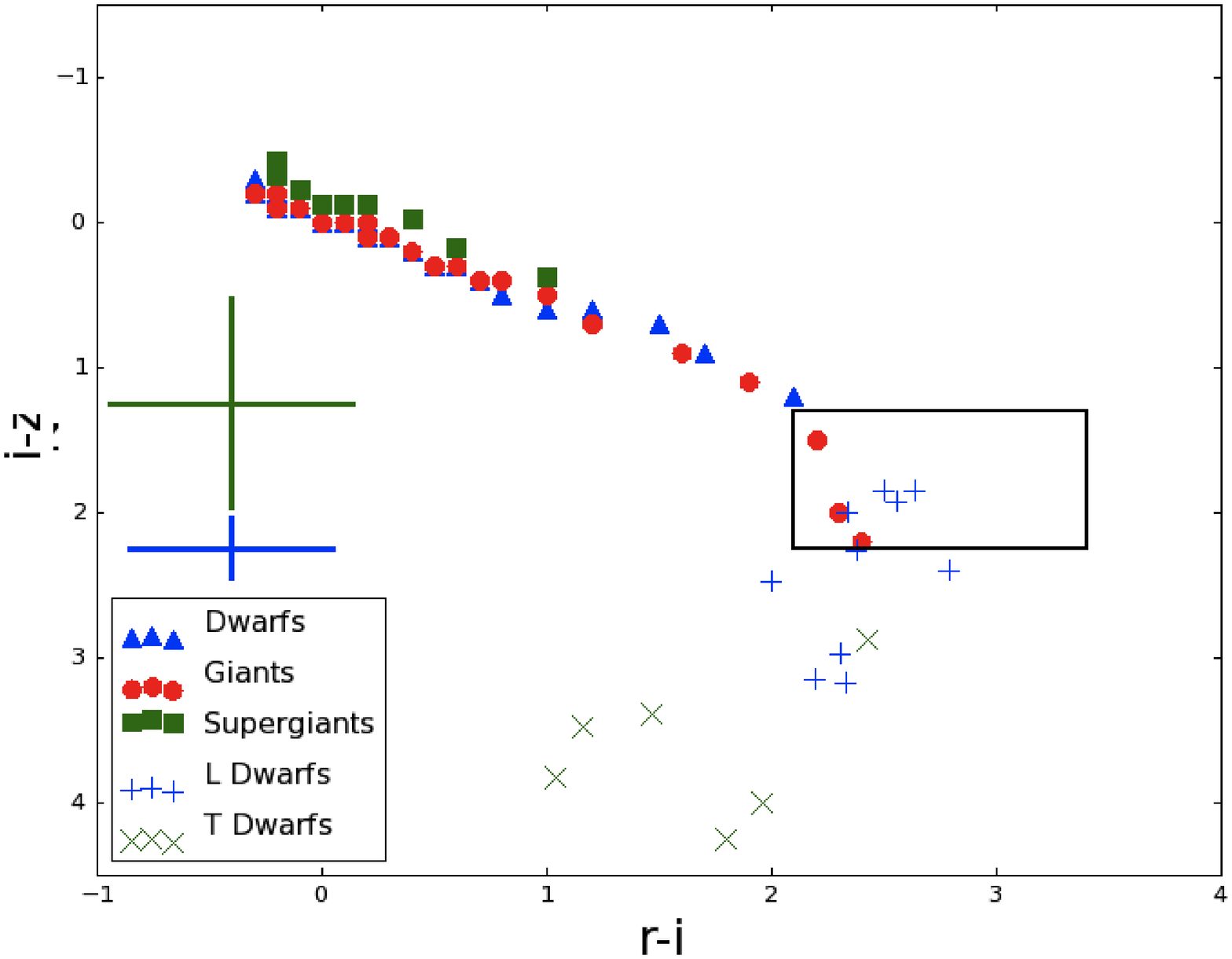} &\includegraphics[width=3.2in]{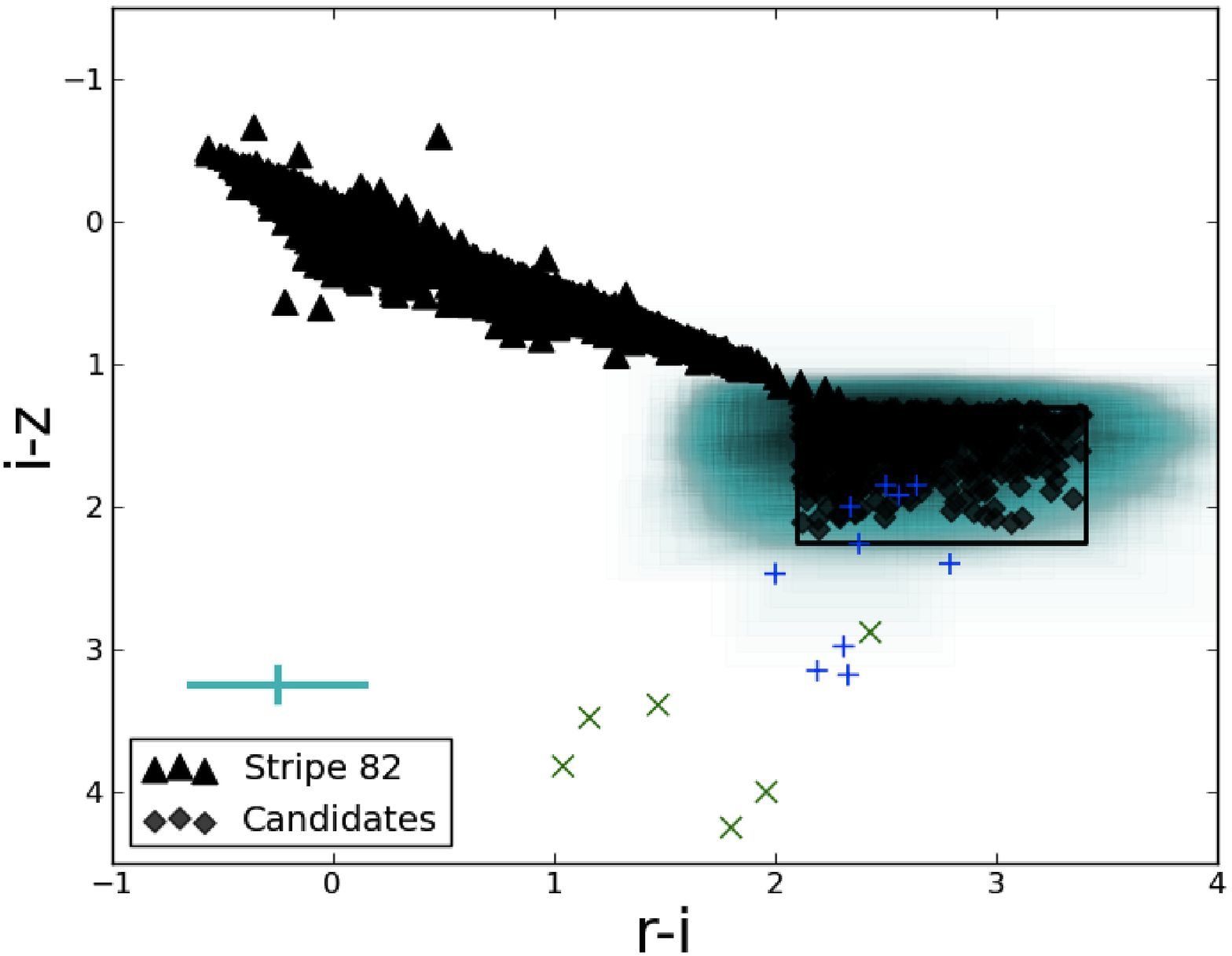}  
\end{tabular}
\caption{Left: Color-color diagram similar to Figure 1, including L and T dwarfs from \protect\cite{Hawley}. The region of color space containing our IGS candidates is marked by the black rectangle. Blue triangles represent dwarfs, green squares represent supergiants, red dots represent giants, blue crosses and green stars represent L and T dwarfs. The blue and green error bars in the bottom left corner are representative of the typical 1-$\sigma$ error bars for L and T dwarfs, respectively. Note that late M (M7-M9) and L dwarfs also contaminate the space. Right: Same as figure on the left, using SDSS Stripe 82 stars (filled triangles) to represent the spread in the stellar locus (the errors are contained within the size of the point) \protect\citep{Ivezic} and 677 extinction-corrected IGS candidates (filled diamonds). The cyan cloud results from the sum of gaussian-distributed errors on each candidate, i.e. the darkest cyan region represents the most probable location of the data. Similarly as in the left panel, the cyan error bar in the bottom left corner is representative of the typical 1-$\sigma$ error bars for each candidate.}
\label{}
\end{figure*}

\section{METHODS}
\label{methods}

During its eight years of operation, the Sloan Digital Sky Survey \citep[SDSS;][]{York} obtained deep, multi-colored images covering more than a quarter of the sky. The SDSS uses 5 optical bandpasses \citep[$\it{u, g, r, i, }$ and $\it{z}$; ][]{Fukugita,Gunn1998,Hogg,Gunn2006} with magnitude limits 22.0, 22.2, 22.2, 21.3, and 20.5, respectively. The DR7 data set contains 12,000 square degrees of images and a catalog of over 350 million objects with spectra of 460,000 stars.

As individual red giant stars in M31 have been observed down to the SDSS magnitude limits \citep[e.g.,][]{Zucker2007}, intragroup red giant stars will be observable. Indeed, at a distance of 300 kpc, all supergiants and roughly half of any giants would be detectable in the $\it{r, i,}$ and $\it{z}$ bands. 

We developed our technique using the synthetic SDSS and 2 Micron All Sky Survey (2MASS) photometry \citep{Covey} of flux calibrated spectra of $\it{solar \ metallicity}$ main sequence, giant, and supergiant standard stars \citep{Pickles}. With simulated SDSS and 2MASS colors \citep{Schlegel}, we obtained a stellar locus to search for giants or supergiants that are isolated in a `clean' region of color space. These color-color diagrams are shown in Figure 1. We choose solar metallicity standards to probe for IGS generated by three-body interactions within the central regions of the Galaxy, as discussed in Section 1.

Unfortunately, most supergiants and giants lie along the main sequence locus, and would therefore be indistinguishable by SDSS colors. However, there is a small area in each of the color-color diagrams, shown with dashed boxes in Figure 1, where the rarest M giants (M7-M10) are isolated from dwarfs and supergiants of the same spectral types. Considering the distances we are probing and the very red colors of these spectral types, we restrict our color selection to the $\it{r, i,}$ and $\it{z}$ bands as follows:
\begin{equation}
2.1 < r - i < 3.4 \text{,} 
\end{equation}
\begin{equation}
1.3 < i - z < 2.2 \text{} 
\end{equation}
Again, we drop the $\it{u-g}$ and $\it{g-r}$ cuts because objects at these distances are likely too faint in these bands.

Since these objects are so red, they may be confused with other nonstellar objects\footnote{Background red galaxies have colors roughly $0.35 < r-i < 0.45$ and $0.15 < i-z < 0.3$ at z=0.1 \citep{Blanton2003}. These colors will become redder with increasing redshift. For example, the brightest and reddest galaxies in SDSS, LRGs at z=0.5, have $r-i=0.7$ and $i-z=0.4$ \citep{Blanton2003}. Since these colors do not approach our color selection region and since we have ensured stellar point spread functions for our candidates, we do not consider background galaxies as a significant source of contamination.}.  However, we find that even quasars with $z>4.6$ are too blue in $\it{i-z}$ to fall in our color space  \citep{Richards,Fan2001}. 

A more worrisome source of contamination comes from L and T dwarfs. To investigate this, we compared our color selection region to the colors of L and T dwarfs \citep{Hawley} and find that they also are contained in the color region, albeit with large uncertainties, as shown in Figure 2. From current observational studies, we estimate that there may be $O(1000)$ early L dwarfs in the SDSS footprint within the magnitude limits of SDSS \citep[][]{Burgasser}. Since we expect more dwarfs than giants at these faint magnitudes, it is likely that  a greater number of dwarfs are scattered into the selection area through large errors than the number of giants scattered out. Objects in our selection box that are not IGS are, nevertheless, likely interesting astrophysical objects yielding more distant L dwarfs than currently known.  We discuss ways to differentiate between IGS and contaminants in section 5.

From the DR7 Star Table, we identified all stars that satisfy our color criteria and are positioned at $|b|>20$ to exclude potential disk contamination (including disk L and T dwarfs). To ensure that each candidate has a stellar point spread function, we confirmed that the object type flag in all 5 bandpasses were stellar\footnote{Although we restrict our color criteria to the $\it{r, i,}$ and $\it{z}$ bands, in order to be conservative, we still require the objects to have stellar point spread functions in the $\it{u}$ and $\it{g}$ bands as well.}. We then removed all objects that would be nearer than 300 kpc and further than 2 Mpc in the redder bandpasses, assuming an M giant average absolute magnitude in each bandpass ($M_{r}=0.8, M_{i}=-1.5, M_{z}=-3.5$) yielding the following magnitude cuts: $23.2<r, 20.9<i, 18.9<z$ --- since the r band is effectively below the magnitude limits of SDSS, we searched for null detections in this band as well. We also eliminated objects that triggered any of the following flags: BADSKY, BLENDED, CHILD, COSMIC\_RAY, EDGE, MAYBE\_CR, MAYBE\_EGHOST, MOVED, NODEBLEND, and PSF\_FLUX\_INTERP.
 
We cross-checked our candidates with the 2MASS J-H color cut of \cite{Majewski} since dwarfs and giants have distinctive J-H colors: $J-K_{s} > 0.85$, $J-H < 0.561(J-K_{s}) + 0.36$, $J-H > 0.561(J-K_{s}) + 0.22$ (top panel of Figure 3).  Although, note that the \cite{Majewski} color cut selects sub-solar metallicity M giants with [Fe/H]$=-0.4\pm{1.1}\ \rm{dex}$  \citep{Chou}. This will make the giants selected by the \cite{Majewski} cut bluer than the solar-metallicity giants selected in our sample.

The middle panel of Figure 3 shows a clear separation of the synthetic spectral standard giants and dwarfs from \cite{Pickles} at a J-H color around 0.8 \citep{BessellBrett}. In general, J, H, and K$_s$ magnitudes for our IGS candidates (16.86, 15.78, and 15.56, respectively, for an M7III according to \cite{Covey}) are too faint to appear in the 2MASS catalog. Since appearing in 2MASS would indicate that the source is too bright or too close to be an IGS M giant we removed any candidate that did appear in the 2MASS catalog, and to verify that they are nearby dwarfs, we plotted their J-H colors shown in the bottom panel of Figure 3.

The completeness limits of J, H, and K filters in the UKIDSS \citep{UKIDDS} survey are appropriate for our targets, however our search through the publicly-released database (DR4) did not result in any matches to our candidates since we expect our candidates lie in parts of the sky not yet covered by this release. Moreover, once the UKIDDS survey is complete it will only cover about half of the area in the SDSS footprint.

After removing known dwarfs in 2MASS, we cross-referenced the IGS candidates with the USNO-B catalog, an all-sky catalog that contains positions, proper motions, and magnitudes in various optical passbands. Here, we removed those candidates with discernible proper motions, ranging from 3 to 1412 mas/year, since the distances determined by the proper motions indicate that they are likely objects closer than 353 pc. This eliminated 782 candidates.

\begin{figure}[]
\centering
\includegraphics[width=0.4\textwidth]{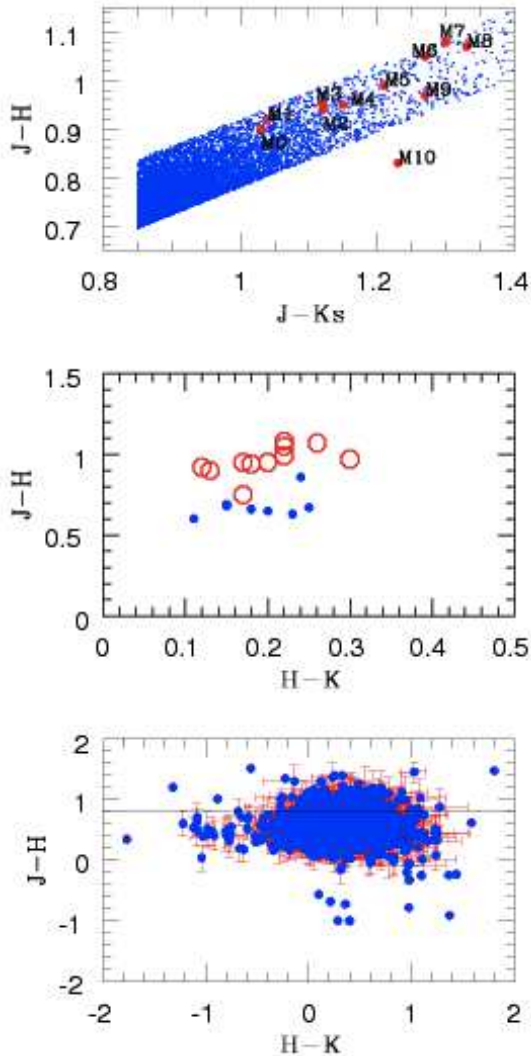}
\caption[]{Top: Blue dots are SDSS Stripe 82 stars \protect\citep{Ivezic} that satisfy the \protect\cite{Majewski} M giant color cuts. Larger red dots are M giants from \protect\cite{Covey}. The point located outside the color cut corresponds to an M10 spectral type. Also note that the late M giants (M7-M10) are located in the part of this color region that is least populated, so they will be least likely to be identified by this cut. Middle: Red open circles are M giants from \protect\cite{Covey}. Blue dots are M dwarfs from \protect\cite{Covey}. Bottom: IGS candidates with 2MASS JHK colors. The solid line represents the J-H = 0.8 separation between dwarfs and giants.}
\label{}
\end{figure}

\subsection{Testing the color selection region with real data}

Since our current technique for finding very late M giant IGS is based on idealized solar metallicity spectra and synthetic colors, it is important to determine how robust these colors are for known M7 III-M10 III stars. Unfortunately, there are no confirmed very late M giant stars in the SDSS DR7 (or DR8) database with both spectra and colors. While it is true that these stars are rare, the real difficulty is that spectral classification of M giants is notoriously difficult, and the latest M giants can be spectral type variables, as well.   Since the SDSS database did not explicitly include the latest M giants, we decided to take a three-pronged approach in checking M giant colors.

First, we searched through Simbad for spectroscopically-confirmed M7-M10 giants, finding 53, 28, 4, and 0, respectively; many of these were listed in the Catalogue of Stellar Spectral Classifications \citep{Skiff2010}. For each object, we cross-checked the spectral type through all other publically-available catalogs on VizieR to determine if the star was a known significant spectral variable, and if so, we discarded it.  We then searched DR7 for any star within 2 arcseconds of the target and obtained the photometry. Many of these objects were nearby and saturated, and therefore appeared as several non-stellar sources in the DR7 database -- these were also discarded if the composite, corrected photometry was not available. Only one of the remaining stars returned a result in the CAS and appears as the blue star in Figure 4, which lies within our color cut.

Second,  we acquired spectra of late-type giants observed as part of a study to quantify the effects of gravity on the spectra of cool objects. These spectra were obtained with the Low Resolution Imaging Spectrometer \citep[LRIS,][]{Oke1995} on the 10-m W. M. Keck Observatory as part of a campaign to construct a systematic surface gravity ``grid" to further constrain spectral classifications of brown dwarfs \citep{Kirkpatrick}. We estimated the spectral type and log g parameters by eye and separated the latest M giant subset for analysis, totaling 13 spectra. Examples of these spectra are shown in Figure 5. Using the SDSS transmission curves, we calculated the colors in the Sloan bands, and as can be seen by the red stars in Figure 4, most of these stars do indeed lie in the predicted color space.

Finally, as a further check of our M giant colors we used the Bruzual-Persson-Gunn-Stryker \citep[BPGS, hereafter;][]{Laidler} stellar atlas of standard stars to obtain synthetic colors using the IRAF Synphot task $\emph{calcphot}$ \footnote{The i-z colors returned by $\emph{calcphot}$ for the M dwarfs are unreliable due to negative flux values in the BPGS stellar atlas.}; the green points in Figure 4 represent these standard stars-- there were no available M9III-M10III standards in the atlas, which is expected because these objects are all spectrum variables.

Figure 4 shows that the color cut we defined from synthetic SDSS colors is consistent with the colors of known M giants from all three techniques. 

If the reader is interested in seeing the colors of the stars we used from the BPGS stellar atlas in the Johnson-Cousins filter set, a figure will be available online \footnote{http://astro.phy.vanderbilt.edu/$\sim$palladl2/}. The UBV colors of these stars are consistent with the colors reported by \cite{Worthey}, as well.

\begin{figure}[]
\centering
\includegraphics[width=0.4\textwidth]{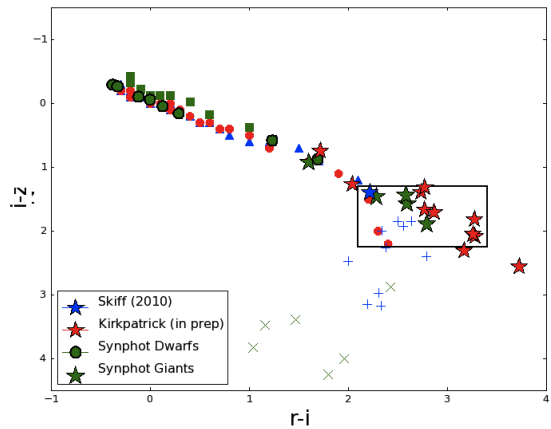}
\caption[]{Same as the left panel of Figure 2. The green stars overplotted here represent M-giant standards, with spectral types between M6III and M8III, from the BPGS stellar atlas. The larger green dots are dwarfs with spectral types O through M, from the same atlas. The photometry for these stars was obtained by implementing IRAF $\emph{synphot}$ tasks. The blue star represents the M giant identified from the Catalogue of Stellar Spectral Classifications \protect\citep{Skiff2010} with confirmed SDSS photometry. The red stars are the M giant contaminants in the \protect\cite{Kirkpatrick} data for which we received spectra.}
\label{}
\end{figure}

\section{RESULTS}
\label{results}

We found a small region of color space, shown in Figure 2, in which the reddest solar-metallicity M giants are isolated from the rest of the stellar locus. This region hosts M7III-M10III stars, along with L dwarfs \citep{Hawley}. 

Using our color selection criteria, we found 159,108 extinction-corrected objects in SDSS DR7. After applying the distance cut and checking the data flags, we narrowed the sample to 4181 objects. We then cross-correlated our sample with the 2MASS and USNO-B surveys, removing any stars with dwarf-like J-H colors and any stars with non-zero proper motions. Our final sample contains 677 IGS candidates. Table 1 lists positions, asinh magnitudes, $\it{r-i}$, and $\it{i-z}$ colors for all 677 candidates. The right panel of Figure 2 shows the location of the final set of IGS candidates with errors in $r-i/i-z$ color space.

\section{Discussion}
\label{discussion}

As discussed in Section 1, IGS could be formed from several different methods. Considering the Local Group's current level of interactions, this population may likely be comprised of high metallicity hypervelocity stars (HVS) ejected through the three body mechanism. However, not enough is known about HVS and Local Group formation to say this definitively, so probing the IGS sample may help us to constrain either or both of these.

If every candidate were a solar metallicity IGS giant, they would be rare tracers of a large underlying IGS population. Assuming a single burst of star formation 10 Gyrs ago and a Salpeter initial mass function (IMF), these candidates represent $O(10^{-4})$ of the total number of IGS and $O(10^{-3})$ of the total mass in IGS, and varies only slightly for differing choices of IMF.

It is useful to compare this to theoretical predictions of stellar ejections from the Milky Way \citep[][]{Kollmeier}. Stars ejected from the galaxy center through three-body interactions with a SMBH will typically have much higher metallicity than stars that were stripped from satellite galaxies originating in the outskirts of a galaxy halo \citep[e.g.,][]{JacobyCiardullo,Kirby}. For example, if we assume that all of the IGS are solar metallicity hypervelocity stars ejected by three-body interactions with a binary black hole consisting of a SMBH and an intermediate mass black hole (IMBH), then the total mass in stellar ejecta will be roughly equal to the mass of the IMBH \citep{Yu,Quinlan}. Given the Milky Way SMBH mass of $4 \times 10^{6} M_{\odot}$ \citep{Ghez}, we would require an IMBH with mass roughly $10^{5}$ M$_{\odot}$ as the companion, independent of initial mass function. This IMBH mass is similar to the mass of the IMBH proposed to be responsible for ejecting stars in the central region of the Galaxy \citep{Lang2011}. This yields several IGS per square degree of sky and roughly tens of red giant hypervelocity IGS in the SDSS footprint. 

Similar back of the envelope calculations suggest that there are $O(1000)$ L dwarfs located in the SDSS footprint, and realizing that late-type dwarfs are more common in general, we anticipate that the majority of our IGS candidates are likely L dwarfs.  If these IGS candidates do turn out to be L dwarfs, then we have identified L-type dwarfs at distances of 100-200 pc, which is up to 4 times farther than currently known \citep{Schmidt}.   

In an attempt to determine if the IGS sample has a distinct spatial distribution, we conducted a set of 2-dimensional K-S tests that compared our candidates with template samples drawn from 3 characteristic distributions: 1) an exponential disk with a scale height of 300 pc to mimic an old stellar population, and a distance cut off of 200 parsec to resemble an L dwarf distribution; 2) a random distribution; 3) and a set of observed hypervelocity stars \citep{Brown2009}. We convolved each template data set with the SDSS footprint and employed the same galactic latitude cut as in our IGS sample. Assuming that these are very cool dwarfs, the IGS sample should exhibit the same distribution on the sky as the exponential disk, but the 2-d K-S test revealed otherwise: the probability that these two samples come from the same underlying distribution is only $10^{-4}$. This is ultimately not surprising since we removed any objects with a measurable proper motion, strongly selecting against L dwarfs within the disk. The second test with a random distribution resulted in an even smaller probability of $10^{-5}$, while the third test with the hypervelocity sample yielded a somewhat higher probability of $10^{-2}$. Figure 6 shows the position of our IGS candidates on the sky compared to the hypervelocity sample used here.

\section{Summary and Conclusions}
\label{summary}

We identified 677 intragroup stellar candidates from the SDSS DR7 using color cuts based on solar metallicity spectral standards. These are extremely red stars with $1.3 < i - z < 2.2$, though the color would shift bluer with lower metallicity. As shown in Figure 2, the M giants in the region are not completely isolated. The latest M dwarfs and early L dwarfs are possible sources of contamination. 

Followup photometric observations of our candidates in the near to mid-infrared wavelengths may differentiate between late dwarfs and M giants. Future followup photometric observations with a 4m class telescope may be promising, albeit impractical. For example, the FLAMINGOS instrument on the NOAO 4m telescope could image all of our candidates with a 113 hour total exposure time in each of the J and H bands, and over 600 hours of total exposure time in the K band for a 10-sigma detection, while this likely would not be sufficient to distinguish M giants from dwarfs. Similarly,  the J, H, and K magnitude limits of NIRI on Gemini are appropriate for our targets, although would require a prohibitively long total exposure time of 2031 hours to achieve a S/N of about 12.

Also, it is possible with long-term photometric followup observations on a 1m class telescope to differentiate between dwarfs and giants based on variability, as late-type M giants tend to be highly variable.

\begin{figure}[]
\centering
\includegraphics[width=0.5\textwidth]{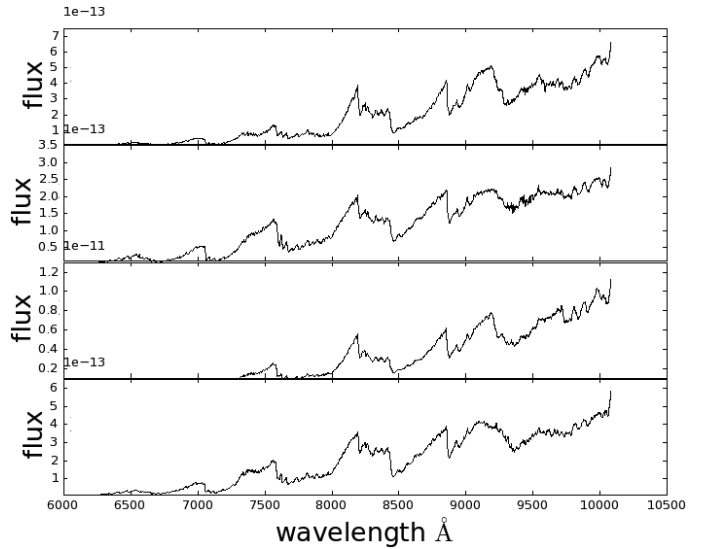}
\caption[\, \,]{Four of the spectra used to compare colors to our IGS candidates. The resulting colors are shown in Figure 4.}
\label{}
\end{figure}

\begin{figure}[]
\centering
\includegraphics[width=0.5\textwidth]{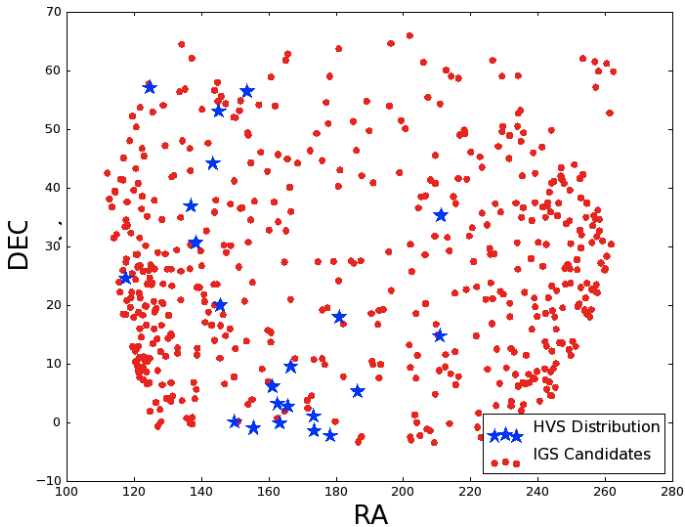}
\caption[\, \,]{The relative positions of our IGS candidates and the hypervelocity stars of \protect\cite{Brown2009} that were compared with the 2-d K-S test. The comparison was made between 522 IGS candidates, red dots, and 22 HVS, blue stars. Notice the higher density of IGS candidates at the edges of the footprint signifying larger numbers of these stars at lower Galactic latitudes.}
\label{}
\end{figure}

\begin{deluxetable*}{ccccccccccccccc}
\tabletypesize{\scriptsize}
\tablecolumns{15}
\tablecaption{IGS candidates remaining after all criteria cuts [Complete version will be available online].}\label{long}
\tablehead{\colhead{Object ID} & \colhead{RA} & \colhead{DEC} & \colhead{u} & \colhead{u err} & \colhead{g} & \colhead{g err} & \colhead{r} & \colhead{r err} & \colhead{i} & \colhead{i err} & \colhead{z} & \colhead{z err} & \colhead{r-i} & \colhead{i-z}}
\startdata
 758882136836343139 & 60.5013 & 80.8760 & 24.4 & 1.3 & 24.7 & 0.6 & 24.3 & 0.7 & 22.0 & 0.2 & 20.3 & 0.2 & 2.3 & 1.7\\ 
 758882137910740030 & 62.9212 & 82.5298 & 25.2 & 1.2 & 25.0 & 0.6 & 23.2 & 0.4 & 21.0 & 0.1 & 19.6 & 0.1 & 2.2 & 1.4\\ 
 758882626993718584 & 63.3170 & 81.6950 & 26.5 & 0.6 & 24.8 & 0.6 & 25.3 & 0.7 & 22.0 & 0.2 & 20.3 & 0.2 & 3.3 & 1.6\\ 
 758877527803168329 & 94.6323 & 63.7038 & 25.1 & 0.9 & 24.1 & 0.3 & 24.5 & 0.5 & 21.2 & 0.1 & 19.8 & 0.1 & 3.3 & 1.4\\ 
 758877527266231607 & 94.8017 & 63.2633 & 24.6 & 0.7 & 24.4 & 0.5 & 24.5 & 0.6 & 21.8 & 0.1 & 20.4 & 0.1 & 2.7 & 1.4\\ 
 758877527266493955 & 95.9369 & 63.5501 & 25.2 & 0.7 & 24.8 & 0.6 & 25.6 & 0.5 & 22.7 & 0.3 & 20.7 & 0.2 & 2.9 & 2.0\\ 
 758878272976455144 & 105.3459 & 66.8553 & 25.3 & 0.9 & 24.4 & 0.4 & 24.1 & 0.4 & 21.8 & 0.1 & 20.2 & 0.1 & 2.3 & 1.6\\ 
 758878271902778853 & 105.8427 & 66.0785 & 23.8 & 0.7 & 24.7 & 0.4 & 24.6 & 0.5 & 22.3 & 0.2 & 20.8 & 0.2 & 2.3 & 1.5\\ 
 758884768580109918 & 107.9936 & 38.3022 & 24.8 & 1.1 & 25.1 & 0.7 & 24.4 & 0.6 & 21.9 & 0.1 & 20.4 & 0.2 & 2.6 & 1.4\\ 
 587738067260998978 & 109.5721 & 39.4395 & 25.3 & 0.9 & 25.7 & 0.6 & 24.6 & 0.7 & 22.1 & 0.2 & 20.0 & 0.1 & 2.5 & 2.1\\ 
\enddata
\end{deluxetable*}

Naturally, low resolution spectroscopic follow-up observations of these IGS candidates would be ideal to confirm their luminosity class. The Calcium II Triplet (CaT) feature at 8498\AA, 8542\AA, and 8662\AA \,is particularly useful for distinguishing late M dwarfs from giants, being much more prominent in the spectra of late-type dwarfs \citep{Reid}. In addition, the strength of the Calcium Hydride (CaH) feature between 6800\AA \,and 7000\AA \,is a good indicator of luminosity class \citep{Cohen}. 

Once the confirmation is complete, we can test the efficiency of our color selection technique, which will be useful for large data surveys like LSST. In fact, two surveys set to launch in the coming year will be particularly well-tuned to find IGS M giants. The DECam survey on the CTIO 4-meter telescope will observe over 1000 square degrees, with magnitude limits of r=23.4, i=24.0, and z=22.9 -- over two magnitudes deeper than SDSS in z. An even deeper survey will launch on the Subaru telescope; the HyperSuprimeCam plans to observe 2000 square degrees down to z=24.9 and y=23.7. Eventually, deeper observations of IGS can reveal the metallicity of these stars -- an important clue to their original birthplace within the group or in situ in the intergalactic medium.

\section{Acknowledgments}
We acknowledge Andreas Berlind for useful suggestions and for allowing us to borrow an SDSS footprint mask code. LEP was supported by GAANN. KHB acknowledges support from NSF Award AST-0847696. JJF acknowledges support from NSF Award AST-0807873.

\bibliographystyle{apj}

\end{document}